\documentclass[aps,twocolumn]{revtex4}
\usepackage{graphicx}
\usepackage{psfrag}
\usepackage{fancyhdr}

\begin{document}
\date{\today}

\title{Pore dynamics of osmotically stressed vesicles}

\author {Yan Levin and Marco A. Idiart}

\address{
  Instituto de F\'{\i}sica, UFRGS, 
  %Universidade Federal do Rio Grande do Sul \\ 
  Caixa Postal 15051, CEP 91501-970, Porto Alegre, RS, Brazil 
}

\begin{abstract}
We present a theory for pore dynamics of osmotically stressed vesicles. 
When a liposome with an internal concentration
of solute is placed inside a solute-depleted medium, an osmotic
flow of solvent through the lipid bilayer leads to
swelling of vesicle and to increase 
in membrane surface tension. This can result in
membrane rupture and opening of thermal pores.  Depending  on the
internal concentration of solute and the size of the vesicle,
pores can close rapidly or be long-lived. We find that the
life span of the long-lived pores scales non-trivially
with the size of the liposome.  Closure of the
long-lived pore is followed by a rapid flicker-like
opening and closing of short-lived pores. Our model is consistent
with the observation of long-lived pores in red blood cell ghosts.

\end{abstract}

\maketitle

Opening of stable pores in cellular phospholipid membranes is an important 
step for  drug delivery and gene therapy. A number of methods have
been explored to this end.  Electroporation~\cite{BeBeZi79} 
is the method of choice
when gene delivery is performed {\it in vitro}. It is, however,
associated with a significant cell mortality limiting its practical use.
Holes in cellular membranes have also been opened by intense
illumination~\cite{BaFrMo95,MoNeBa97}.  
The mechanism leading to pore formation under these
circumstances is still not fully understood. It, however,
appears to be similar to the action of laser tweezers~\cite{Gr00},
a part of the low dielectric constant
lipid membrane is sucked into the laser trap, stretching the  
membrane until rupture.

For drug delivery the goal is to introduce 
a controlled amount of substance
to the specific disease site. The delivery system
should minimize the potential side-effects  by selectively targeting
the affected region of the organism~\cite{MaFeCu01}. 
Liposomes can be made to 
contain  drugs or genes and can be introduced into 
blood stream without provoking an immune system response.  
Liposomal affinity 
for specific tissue can be increased by varying the membrane lipidic 
composition or by including  ligands 
that recognize and bind to specific receptors~\cite{Pa02}.
One example of this is the drug called  
Doxirubicin used in cancer chemotherapy.  The liposomal encapsulation of
Doxirubicin has been shown to efficiently deliver it to tumors, while
minimizing the side effects, such as damage to heart muscle tissue.  

When liposome ruptures, its content leaks out.  It
is precisely the dynamics of this rupture that is the subject
of the present  Letter. In particular, our goal is to understand 
the mechanism of pore opening and closure in the 
osmotically stressed vesicles.
The situation which we attempt to model is the following. Liposomes
are prepared inside a solution containing small molecules 
impermeable to lipid membrane
at concentration $c_0$. They are then
placed into a purified aqueous solution.  The imbalance in solute
concentration inside and outside the vesicles leads to
osmotic flow of solvent through the semi-permeable lipid membrane,
resulting in swelling of vesicles and stretching of their membranes. 
As the tension increases, the energy barrier to 
pore nucleation decreases~\cite{DeGu62}.  If there is enough solute 
inside the liposome, the barrier height
eventually becomes comparable to the energy of thermal fluctuations.  
When this happens, membrane  raptures
and a  pore opens~\cite{TaDvSa75}.  
One interesting question is what is the life
span of this pore?  Specifically, as a pore opens it relaxes the membrane
tension.  However, opening of a pore exposes the hydrophobic
interior of membrane to water, leading to pore line (edge) tension.
This hydrophobic effect induces a force which tries to re-seal the 
pore.  Furthermore, a hole in membrane results in  
efflux of both solute and  solvent,
reducing the vesicle volume and surface tension.
All these effects lead to an intricate interplay of time 
and length scales
which control the pore closure.

We shall designate the radius of a vesicle as $R$ and the radius of a
pore as $r$.  The conservation of mass then leads to
%-----------------------------
\begin{equation}
\label{1} 
4 \pi \rho R^2 \frac{d R}{dt}=j_w-\pi r^2 \rho v\;,
\end{equation}
%-----------------------------
where $\rho$ is the density of water, $j_w$ is the osmotic flux
and $v$ is the leak-out velocity. The  leak-out 
velocity~\cite{HaBr86,ZhNe93,SaMoBr99,BrGeSa00} 
is determined by the balance between the
shear stress,  proportional to $\eta v/r$, and the osmotic 
pressure inside the vesicle $\Delta p$,
%-----------------------------
\begin{equation}
\label{2} 
v=\frac{\Delta p r}{3 \pi \eta}\;,
\end{equation}
%-----------------------------  
where $\eta$ is the solvent viscosity. Inside a swollen
vesicle the osmotic pressure is compensated by the Laplace pressure,
so that 
%-----------------------------
\begin{equation}
\label{2a} 
\Delta p=\frac{2 \sigma}{R}\;,
\end{equation}
%-----------------------------
where $\sigma$ is the tension of
a stretched membrane. In this work we shall concentrate on vesicles
with fairly large internal concentrations of solute, close to
$1$ M.  This corresponds to osmotic pressures as high as
$20$ atmospheres.  Of course, before these extreme  pressures
can be built, the lipid membrane will 
rapture, leaking out some of the internal content and releasing the
surface stress. At these high 
pressures, the thermal undulations of  membrane can be ignored,
and the membrane tension is 
controlled by the stretching modulus~\cite{HeSe84,EvRa90,SeIs02} $K_s$.
The  membrane elastic energy takes a Hook-like form 
%-----------------------------
\begin{equation}
\label{3} 
E_s=\frac{1}{2}K_s (A-A_0)^2\;,
\end{equation}
%-----------------------------
where $A_0=4 \pi R_0^2$ is the equilibrium surface area of the 
unstretched membrane. As the membrane is stretched beyond the
elastic limit, pores are nucleated reducing
the membrane tension. The elastic energy of a membrane 
with a pore is
%-----------------------------
\begin{equation}
\label{4} 
E(R,r)=\frac{1}{2}K_s [4 \pi (R^2-R_0^2)-\pi r^2]^2+2 \pi \gamma r\;,
\end{equation}
%-----------------------------
where $r$ is the pore radius and $\gamma$ is the pore line tension. 
The membrane  surface tension is,
%-----------------------------
\begin{equation}
\label{5} 
\sigma(R,r)=\frac{\partial E_s}{\partial A}=
K_s (4 \pi (R^2-R_0^2)-\pi r^2)\;.
\end{equation}
%-----------------------------

The osmotic current $j_w$ is determined by 
the membrane permeability $P$, the
concentration difference of solute inside and outside the vesicle,
and the Laplace pressure. 
A simple phenomenological expression for the osmotic current
of water into the vesicle can be written as,
%-----------------------------
\begin{equation}
\label{7} 
j_w=P (4 \pi R^2-\pi r^2)\left[c - \frac{2 \sigma}{10^3 k_B T N_A R}\right]\;,
\end{equation}
%-----------------------------
where $k_B$ is the Boltzmann constant and  $N_A$ is the Avogadro number.
If the difference of molar $(M)$ solute concentration $c$, 
inside and outside
the vesicle is not very large, the integrity of membrane
will not be compromised,  and a stationary state with
$j_w=0$ will be achieved. Under these conditions the osmotic pressure
is completely compensated by the Laplace pressure, 
resulting in zero net flux
of solvent.  For sufficiently {\it large} internal concentration $c_0$, 
a stationary
state will {\it not} be achieved before membrane ruptures.  It is
precisely in this regime that we expect to see some interesting physics.

The growth of a pore is controlled by the rate at which the elastic energy
is dissipated by the  membrane viscosity $\eta_m$~\cite{DeMaBr95},
%-----------------------------
\begin{equation}
\label{6} 
\eta_m l\frac{d r}{d t}= \sigma(R,r) r -\gamma\;,
\end{equation}
%-----------------------------
where $l$ is the membrane width.
Since the membrane
is impermeable to solute particles the internal solute
concentration
is modified only through the osmotic influx of solvent 
and the efflux of solute through the open pore, after 
membrane has ruptured.
The continuity equation expressing this is  
%-----------------------------
\begin{equation}
\label{8} 
\frac{4 \pi}{3} R^3 \frac{d c}{d t}=-4 \pi R^2 c \frac{d R}{d t}- 
\pi r^2 c v\;,
\end{equation}
%-----------------------------
where we have assumed that solute is uniformly distributed inside
the vesicle.

We are now in position to study the evolutionary dynamics of an osmotically
stressed vesicle. The characteristic values of the
parameters appearing in the above equations are:
$4 \pi R_0^2 K_s=0.2\, J/m^2$, $l=3.5 nm$, 
$\gamma=10^{-12}\,J/m$, $\eta_m=100\, Pa s$ ,
$\eta_w=10^{-3}\, Pa\,s$  and $P=1.8 \times 10^{-4}\, kg/m^2\, s\, M$.

The energy necessary to open a 
pore or radius $r$ is $\Delta E(R,r)=E(R,r)-E(R,0)$.  The form of  
$\Delta E(R,r)$ as a function of pore size is plotted in Fig. 1
for various values of $R$. 
%%%%%%%%%%%%%%%% figure %%%%%%%%%%%%%%%%%%%%%
\begin{figure}[h]
\begin{center}
\includegraphics[angle=270,width=6cm]{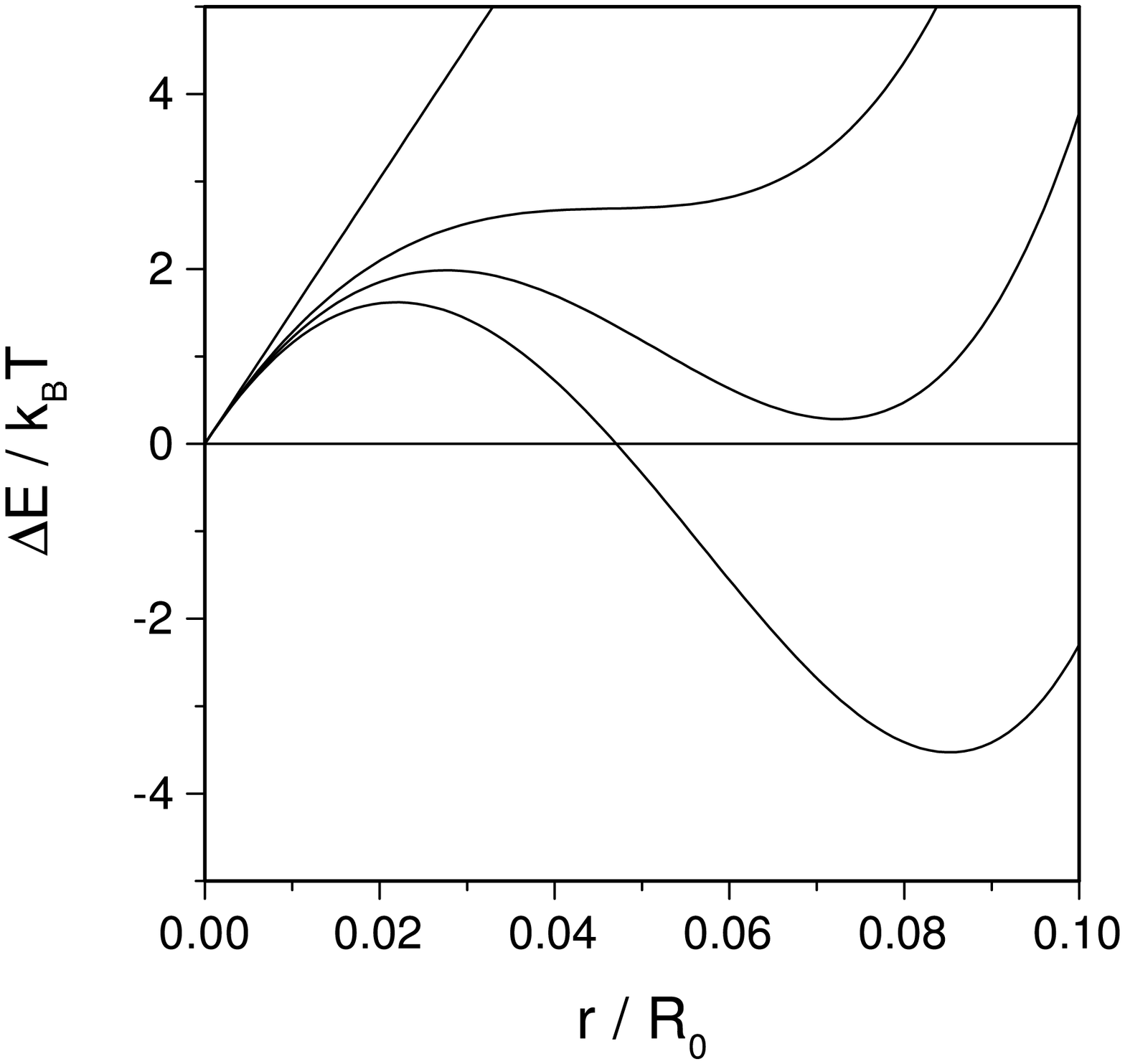}
\end{center}
%\vspace{1cm}
\caption{Energy $\Delta E(R,r)$ necessary to open a pore of 
radius $r$ in a liposome with $R_0=100 \,nm$. 
The curves are, from top down,  $R/R_0=1.0$, $R/R_0=1.0008$,
$R/R_0=1.001$, and $R/R_0=1.0012$ }
\label{figure1}
\end{figure}
%%%%%%%%%%%%% end of figure %%%%%%%%%%%%%%%%% 
As the osmotic flow of solvent
stretches the membrane, a minimum in $\Delta E(R,r)$ for $r \ne 0$
indicates
energetic favorability of pore formation. However, to reach this
minimum  a sufficiently large thermal fluctuation is 
necessary to overcome the energy barrier. 
The probability of such a fluctuation occurring is proportional
to the Boltzmann factor
%-----------------------------
\begin{equation}
\label{9} 
P(r) \sim e^{-\beta \Delta E(R,r^*)}\;,
\end{equation}
%-----------------------------
where $\beta=1/k_B T$ and  $r^*$ is the critical pore size.
Pores with $r<r^*$ will shrink and close while pores with
$r \ge r^*$ will grow. The waiting time for 
appearance of a thermal fluctuation with enough energy to 
open a pore of radius $r^*$ is very long, 
unless $ \Delta E(R,r^*)\approx k_B T$.
 
In the limit of large surface tensions the
critical pore size is $r^*\approx \gamma/\sigma$ 
and the barrier height is
%-----------------------------
\begin{equation}
\label{10} 
\beta \Delta E_b \approx \frac{\pi \gamma^2}{\sigma k_B T} \;.
\end{equation}
%----------------------------- 
We note that when  membrane tension 
reaches $\sigma_c=10^{-3}\, J/m^2$,
the barrier height is approximately $k_B T$. 
This value of $\sigma_c$ is in agreement with the tensions
found to be necessary to rupture a mechanically 
stretched membrane~\cite{Ti74,WeCh96}.
The minimum solute concentration necessary for a  pore to open is
%-----------------------------
\begin{equation}
\label{11} 
c_0^{min} \approx \frac{2 \sigma_c}{10^3 k_B T N_A R_0}\;,
\end{equation}
%----------------------------- 
which for a vesicle of $100$ nm is approximately $10 \,mM$. 

The dynamics then proceeds as follows: 
Eqs. (\ref{1}),(\ref{6}), and (\ref{8}) are 
solved numerically, using Euler's method with a stepsize of 
$\delta t=0.1\mu s$,
to find the evolution of $R(t)$, $r(t)$, and $c(t)$.  
When the  surface tension becomes such that, 
$\Delta E_b \approx k_BT$,
a pore  of size $r^*$ opens, see Fig. 2, and the
internal content of the liposome begins to leak-out.  
This process reduces the volume of the vesicle and decreases the membrane
tension, until the pore closes. Once this happens, the 
osmotic swelling of the liposome re-starts.  
As the energy barrier to pore nucleation 
drops down to $k_B T$ a new pore is opened, etc.
The process  stops when the internal solute 
concentration reaches $c_0^{min}$ and a stationary state with
$j_w=0$ is established.
%%%%%%%%%%%%%%%% figure %%%%%%%%%%%%%%%%%%%%%
\begin{figure}[h]
\begin{center}
\includegraphics[angle=270,width=8cm]{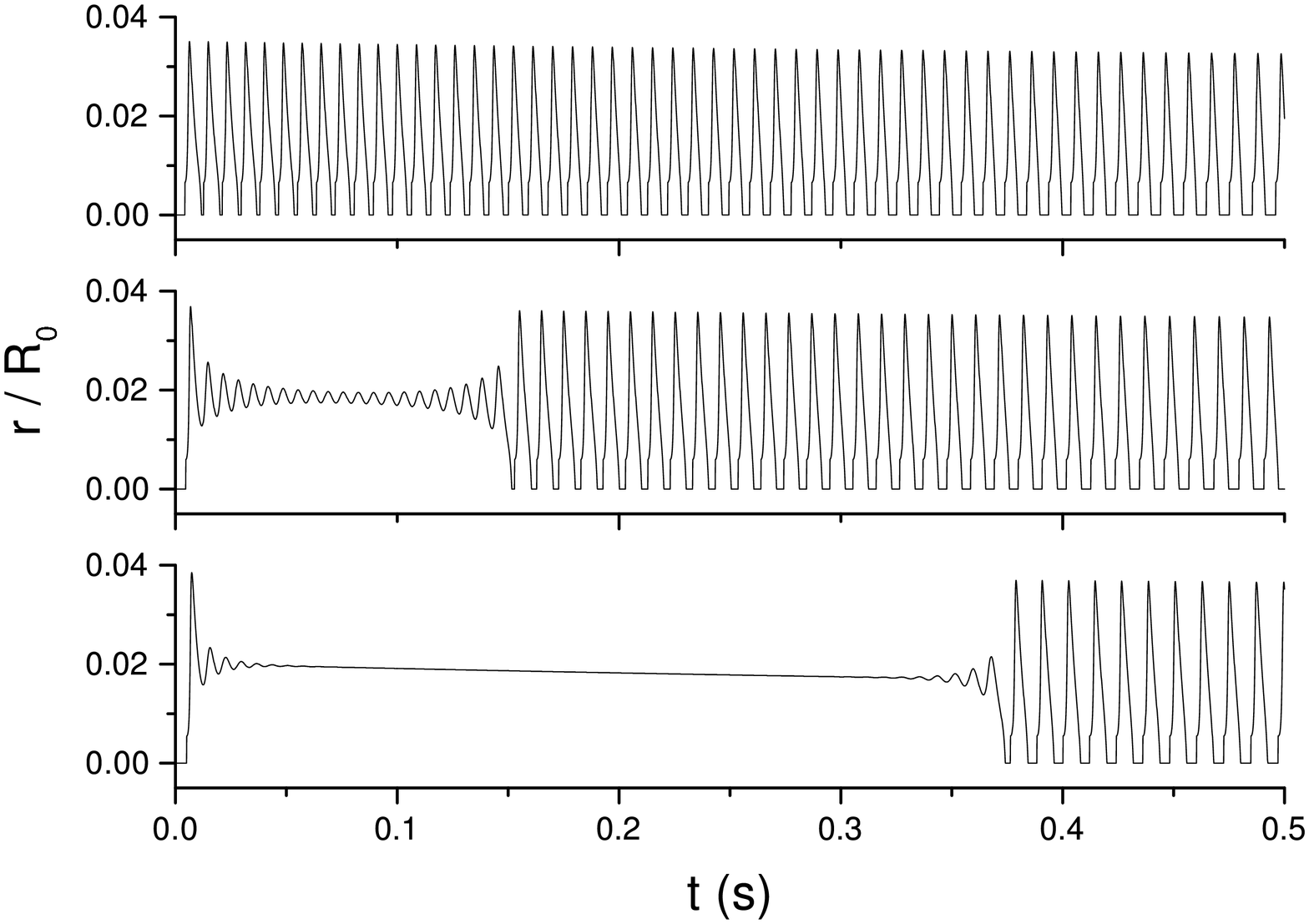}
\end{center}
%\vspace{1cm}
\caption{Radius of a pore as a function of time for vesicles
of $R_0=200\, nm$ (topmost), $220 \,nm$ and $240\, nm$ (bottommost),
for initial concentration  $c_0 = 0.5 M$.
Note that for vesicle with $R_0=200 \,nm$ the pores are short-lived,
while for larger vesicles, long-lived pore opens first.}
\label{figure2}
\end{figure}
%%%%%%%%%%%%% end of figure %%%%%%%%%%%%%%%%% 

For small vesicles, pores open and close very quickly with the
characteristic time  $\tau_f \sim 10^{-2} $,
resulting in a periodic flickering.  For liposomes
with $R_0>R_0^c(c_0)$ a long-lived pore appears, see Fig. 3.  
The critical size of a liposome $R_0^c(c_0)$, necessary for
nucleating a long-lived pore depends on the
internal solute concentration. Larger the initial solute concentration $c_0$, 
smaller will be the size of the vesicle which supports a
long-lived pore.   The long life
span of these pores is the result of a ``wash-out'' effect in which
the osmotic flux is almost completely compensated by the leak-out rate 
of solute through the pore. When solute
concentration inside the vesicle drops below the critical value,
$c_c(R_0)$,  the long-lived pore closes. This value is 
insensitive to  the initial solute concentration $c_0$, 
but depends strongly on the 
vesicle size $R_0$.
The life span of a long-lived pore $\tau$ 
scales with the vesicle size $R_0$,
%-----------------------------
\begin{equation}
\label{12} 
\tau \sim R_0^\nu\;,
\end{equation}
%-----------------------------
with $\nu \approx 2.3-2.4$, see Fig. 3.
After the long-lived pore has closed, it is followed by a sequence
of short-lived pores with the characteristic life span $\tau_f$. 
%%%%%%%%%%%%%%%% figure %%%%%%%%%%%%%%%%%%%%%
\begin{figure}[h]
\begin{center}
\vspace{0.5cm}
\includegraphics[angle=270,width=8cm]{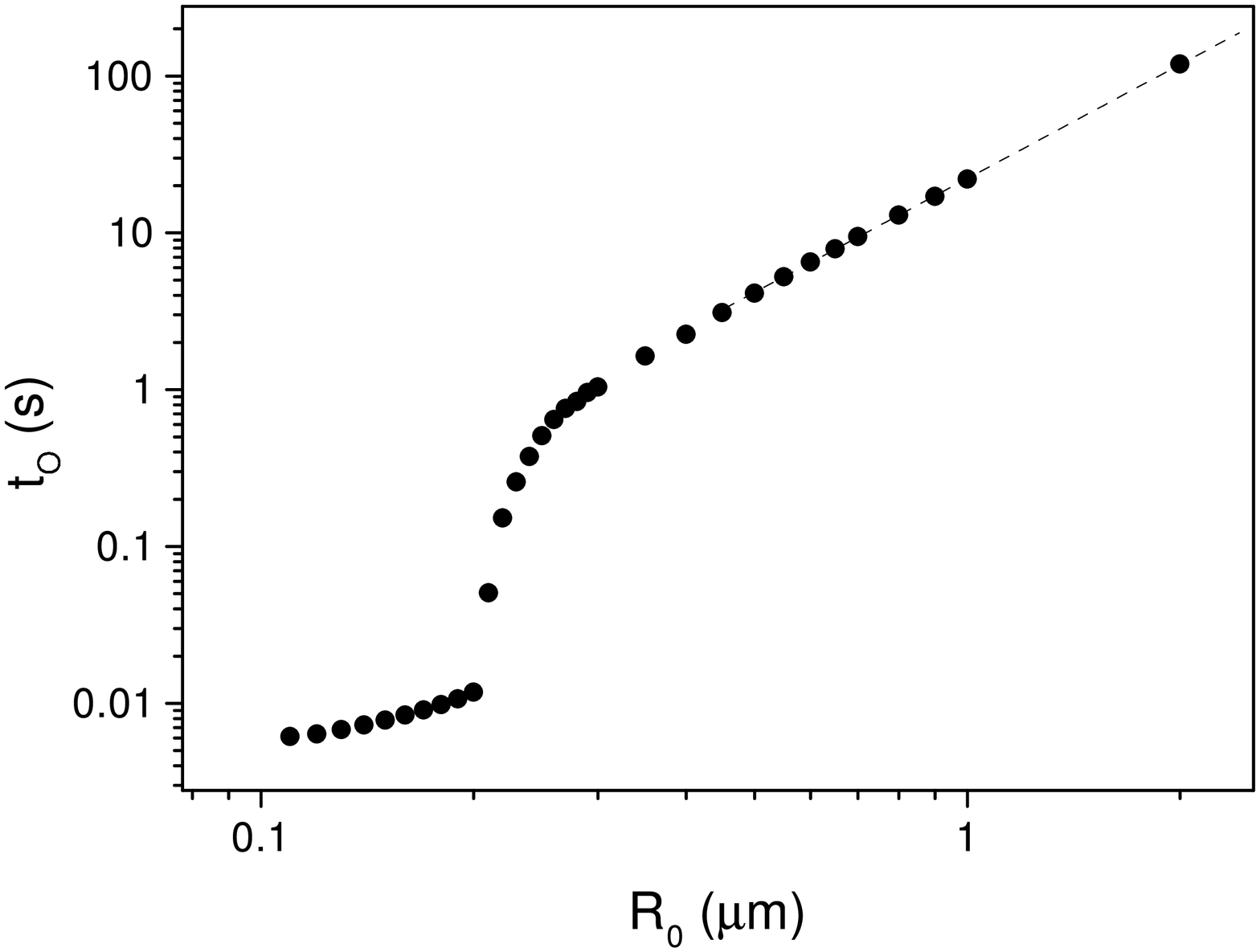}
\end{center}
%\vspace{1cm}
\caption{Life span of the first-open pore as a function as a function of 
vesicle size $R_0$ for $c_0=0.5 M$.  Note the appearance of the critical vesicle
size $R_0^c(c_0)$ which can sustain long-lived pores.}
\label{figure3}
\end{figure}
%%%%%%%%%%%%% end of figure %%%%%%%%%%%%%%%%% 

We have checked that the pore dynamics is not sensitive to the
specific mechanism of pore opening.  In particular,
even if the pores are opened stochastically, with the probability
given by the Boltzmann distribution, the rate of leakage and the
life span of the long-lived pores is affected very little. 

Up to now we have not taken into account the diffusive efflux of 
solute through the open pores. The diffusive current through
a pore of size $r$ is approximately 
%-----------------------------
\begin{equation}
\label{13} 
j_D \sim \pi r^2 c \frac{D}{R_0}\;.
\end{equation}
%-----------------------------
This leads to the decline of internal solute concentration 
governed  by the conservation equation,
%-----------------------------
\begin{equation}
\label{14} 
\frac{4 \pi}{3} R_0^3 \frac{d c}{d t}=-j_D\;,
\end{equation}
%-----------------------------
the solution of which is
%-----------------------------
\begin{equation}
\label{14a} 
c(t)=c_0 e^{-\frac{t}{\tau_e}}\;,
\end{equation}
%-----------------------------
where the effusion time $\tau_e$ is
%-----------------------------
\begin{equation}
\label{15} 
\tau_e=\frac{4 R_0^4}{3 r^2 D}\;.
\end{equation}
%-----------------------------
Using $D \approx 10^{-9}\, m^2/s$, appropriate for small organic molecules such
as sucrose,
and $r=r^*\approx \gamma/\sigma_c \approx 1 \,nm$, we see that for liposomes 
with $R_0=50 \,nm$, the time for effusion is 
$\tau_e \approx 10^{-2}$ s.  This is comparable to
the flicker time $\tau_f$. Therefore, 
for small vesicles effusion is 
an important mechanism for the loss of solute.  On
the other hand, for liposomes with $R_0=100 \,nm$ and 
above, effusion is only marginally relevant.
   
Long lived pores have been observed in
red blood cell ghosts~\cite{St70,StKa74}, their size dependent on the 
ionic strength of the surrounding medium~\cite{LiSt82}. No theory,
up to date, was able to account for these long-lived pores.
Holes were predicted to either grow indefinitely,
which would result in ghost vesiculation, or to 
close completely~\cite{BeBr99}.  
Our model 
provides a dynamical mechanism for pore stabilization, consistent
with the experimental observations. In the specific case of
red blood cell ghosts the ratio $\gamma/K_s$ has to
be ajusted to account for the large radius $r/R_0$ observed in these
experiments~\cite{St70,StKa74}.

In aqueous solutions the phospholipid membranes
acquire a net negative charge. At physiological
concentrations, $154\,mM$ of $NaCl$, the Debye length, however, is
quite short, less then $1\,nm$ 
and the electrostatic interactions are strongly
screened~\cite{Le02}.  We, therefore, do not expect that electrostatics 
will significantly modify the basic conclusions  of our
theory, beyond the renormalization of membrane line~\cite{BeBr99} 
and surface tension.  However, further, investigations in this direction
are necessary and will be the subject of future work.     
 
Finally, it is curious to note a strong similarity between
the rupture of osmotically stressed liposomes and
bursting of {\it Hydra} cells aggregates~\cite{MoAlTh01}. {\it Hydra} 
and marine sponges can generate functional organisms
from random cell aggregates purely through the intercellular
interaction. This morphogenesis is characterized by the
cavity formation followed by swelling 
and violent bursting, which expels the 
internal fluid and dead cells, see  Fig. 5. 

%%%%%%%%%%%%%%%% figure %%%%%%%%%%%%%%%%%%%%%
\begin{figure}[h]
\begin{center}
\vspace{0.5cm}
\includegraphics[angle=270, width=8cm]{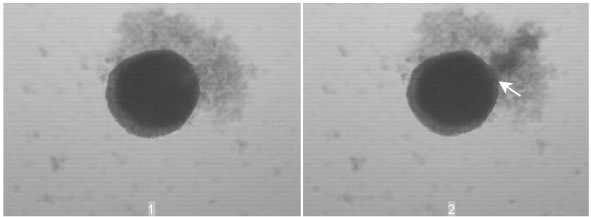}
\end{center}
%\vspace{1cm}
\caption{A sequence of bursts in hydra aggregates. Picture
one shows an aggregate with a diffuse cloud of expelled
cells from the previous explosions, while picture two shows
a cellular aggregate in the process of bursting. Arrow indicates
the site of the burst.}
\label{figure4}
\end{figure}
%%%%%%%%%%%%% end of figure %%%%%%%%%%%%%%%%%
\vspace{4cm}
Unlike the liposomes, cellular aggregates are too massive for temperature
to be of any relevance. Instead the cellular Brownian motion is driven by 
the metabolic fluctuations of the cytoskeleton~\cite{MoGl96}. 
We can,
therefore, expect that a theory similar to the one presented above
for the liposomes, might also apply to osmotically driven
bursting of cellular aggregates.

The authors are grateful to Ms. V. A. Grieneisen for kindly
providing the photos of bursts in hydra aggregates, Fig. 4.
This work was supported in part by the Brazilian agencies
%Conselho Nacional de Desenvolvimento Cient{\'\i}fico e Tecnol{\'o}gico 
CNPq and FAPERGS.    

%\bibliographystyle{prsty}
%\bibliography{references}

\end{document}